# Evolution of variable range hopping in strongly localized two dimensional electron gas at NdAlO$_3$/SrTiO$_3$ (100) heterointerfaces

A. Annadi,[1,2] A. Putra,[1,3] A. Srivastava,[1,2*] X. Wang,[1,2*] Z. Huang,[1,2*] Z.Q. Liu,[1,2*] T. Venkatesan,[1,2,3] and Ariando[1,2,a]

*[1]NUSNNI-Nanocore, National University of Singapore, Singapore 117411, Singapore*

*[2]Department of Physics, National University of Singapore, Singapore 117542, Singapore*

*[3]Department of Electrical and Computer Engineering, National University of Singapore, Singapore 117576, Singapore*

[a)]ariando@nus.edu.sg

**We report evolution of the two-dimensional electron gas behavior at the NdAlO$_3$/SrTiO$_3$ heterointerfaces with varying thicknesses of the NdAlO$_3$ overlayer. The samples with a thicker NdAlO$_3$ show strong localizations at low temperatures and the degree of localization is found to increase with the NdAlO$_3$ thickness. The $T^{-1/3}$ temperature dependence of the sheet resistance at low temperatures and the magnetoresistance study reveal that the conduction is governed by a two-dimensional variable range hopping mechanism in this strong localized regime. We attribute this thickness dependence of the transport properties of the NdAlO$_3$/SrTiO$_3$ interfaces to the interface strain induced by the overlayers.**

The discovery of the two dimensional electron gas (2DEG) at the LaAlO$_3$/SrTiO$_3$ oxide hetero interface[1] has made this material system a primary candidate for applications in oxide electronics. This system has further triggered the interest of the community as novel phenomena, such as superconductivity and magnetism,[2-5] were observed at this interface and furthermore these properties were shown to be sensitive to the external parameters such as O$_2$ deposition pressure[6] and strain effects[7] at the interface. One of the characteristic features of the 2DEG is the overlayer thickness dependence[8,9] of its conductivity, where a thickness dependent metal to insulator transition (MIT) transition was observed at a critical thickness of LaAlO$_3$ overlayer. Furthermore, a weak to strong localization[10,11] and magnetic Kondo behavior[6] were reported in thicker samples, suggesting that a variety of conduction mechanisms were operating at these interfaces. Moreover the superconductivity at these interfaces was also shown to be sensitive to the thickness[2] of the over layer where the localization is suggested to be minimum. Very recently, we reported the formation of 2DEG at various polar/nonpolar oxide interfaces[12] showing a stronger localization of 2DEG compared to that at the LaAlO$_3$/SrTiO$_3$ interfaces. This stronger localization can possibly give rise to different transport properties, and thus its dependence on the overlayer thickness is very crucial to be investigated.

In this paper we investigate the transport properties of the 2DEG formed at the NdAlO$_3$/SrTiO$_3$ heterointerface with NdAlO$_3$ overlayer thicknesses of 6, 12 and 16 unit cells (u.c) grown on TiO$_2$ terminated[13] SrTiO$_3$ (100) substrates by pulsed Laser deposition (PLD). The basic difference between NdAlO$_3$/SrTiO$_3$ and LaAlO$_3$/SrTiO$_3$ interfaces is that the former one offers larger lattice mismatch of 3.6%. The parameters used for PLD growth process are as follows: 780 $^o$C substrate temperature, oxygen pressure of 1x10$^{-3}$ Torr and energy density of the laser ($\lambda$ = 248 nm) of 1.3-1.4 Jcm$^{-2}$, with a pulse repetition rate of 1 Hz. During the film growth, the thickness of the NdAlO$_3$ is monitored by in-situ Reflection High Energy Electron Diffraction (RHEED) as shown in Fig. 1a. Figure 1b shows the Atomic Force Microscopy (AFM) topography image after the deposition of NdAlO$_3$/SrTiO$_3$ sample showing well defined step flow surfaces. The RHEED oscillations and the preserved step and terrace structure shown by AFM imply the layer-by-layer growth of the overlayers. The electrical connections to the heterointerface are made by Al wire bonding and the electrical measurements were performed by Quantum Design Physical Properties Measurement System (PPMS). Sheet resistance ($R_s$), carrier density ($n_s$) and mobility ($\mu$) were obtained by using the Van der Pauw measurement geometry.

Figure 2a shows the $R_s$ variation with temperature for NdAlO$_3$/SrTiO$_3$ interfaces of various thicknesses. The 6 u.c sample shows a typical metallic behavior (albeit a weak upturn at low temperature), whereas thicker samples (12 and 16 u.c) show a metallic behavior (300-70 K) but with strong upturns in $R_s$ with resistance minima at about 70 K. The emergence of the upturns with overlayer thickness observed here is in agreement with previous reports on LaAlO$_3$/SrTiO$_3$.[9] However, a significant observation can be made for the low temperature transport for the thicker NdAlO$_3$/SrTiO$_3$ samples, *i.e.* below 15 K the variation in $R_s$ is more distinct indicating a different type of transport mechanism operating at low temperatures. Figure 2b shows the temperature dependence of $n_s$ and $\mu$ for the corresponding samples. For the thicker samples the $n_s$ decreases in the temperature range of 300-20 K, can be attributed to the localization of activated charge carriers at the interface. The $n_s$ in this range can be fit with a simple Arrhenius equation of the form $Ae^{-E_a/T}$, where $E_a$ is the thermal activation energy for the activated charge carriers. The $E_a$ is of the order of 12 meV which is larger compared to 6 meV for the LaAlO$_3$/SrTiO$_3$ interfaces.[14] The large activation and strong localizations observed in thicker samples would be due to the large structural distortions present in these systems arising from the larger interface strain present. Further, there is an increase in $n_s$ at low temperatures below 20 K down to 2 K, while simultaneously an increase in $R_s$ is observed. For the temperatures below 20 K the increase in both the $R_s$ and $n_s$ implies that the transport mechanism cannot only be simple band (Drude) conduction and indicates that other type of transport mechanism is in operation at low temperatures. Contrary to this, for the 6 u.c sample there is no such carrier activation for $n_s$ with temperature. Meanwhile the $\mu$ for the 6 u.c sample increases monotonically from 9 $cm^2V^{-1}s^{-1}$ at 300 K to 900 $cm^2V^{-1}s^{-1}$ at 2 K. The relatively high mobility of the 6 u.c sample would be favorable for the further studies of superconductivity. On the other hand, for the thicker samples (12 and 16 u.c) the $\mu$ at 300 K is about 5 $cm^2V^{-1}s^{-1}$ and it drops significantly towards the low temperatures, which is further evidence for strong localization in these samples. The low temperature behavior of $R_s$, $n_s$ and $\mu$ confirms the strong localization of carriers associated with a change of transport mechanism at low temperatures.

In general, the observed low temperature sharp upturns in $R_s$ can arise from various transport mechanisms namely, Variable Range Hopping (VRH) due to strong localizations, a magnetic Kondo scattering and thermal activation. In order to further verify the transport mechanism in our system, the low temperature variation of $R_s$ is analyzed. The observed strong localization of 2DEG at the interface suggests that the most likely mechanism would be the VRH, which is a transport mechanism in which the conduction takes place through a hopping of carriers between localized states. The variation of resistance in the VRH regime can be described by the equation[15] $R(T)=R_0 \exp(T_0/T)^{1/(n+1)}$, here $n$ is the dimensionality of the system. Figure 3 shows the corresponding 2D VRH ($n = 2$) fit for the experimental data at low temperatures (40-2 K).

Clearly, the experimental data below 20 K fits well with the VRH formula and moreover the $T^{-1/3}$ dependence ($n = 2$) confirms the 2D nature of the conducting channel. The crossover regime between the metallic and VRH (70-20 K) follows roughly a logarithmic temperature dependence, indicating a weak localization behaviour. On the other hand, the upturns in $R_s$ could originate from magnetic Kondo effect following the dependence given by $R(T)=R_0 \ln (T/T_{ef})$. However this dependence does not fit our data, indicating that Kondo scattering is not the governing mechanism here. Moreover the non saturating trend of $R_s$ even up to 2 K also further confirms this. For the thicker samples the $n_s$ at 20 K are of the order of $\sim 10^{12}$ cm$^{-2}$ which is one order of magnitude lower than that in the thin 6 u.c sample (Fig . 2b). Hence, it seems that there could be a critical $n_s$ at these interfaces below which the transport mechanism converts to VRH in the strong localization regime. The reduced $n_s$ seen in thicker samples could be due to the large amount of interface strain created by thicker NdAlO$_3$ over layers.

The VRH transport mechanism can be further evaluated by magneto transport measurements, since in the hopping transport the carrier conduction is quite sensitive to magnetic fields. Magneto-resistance (MR) measurements performed on the 12 u.c NdAlO$_3$/SrTiO$_3$ sample at various temperatures with different current to magnetic field orientations are shown in Figure 4a which shows the out of plane MR (magnetic field is perpendicular to the current and sample surface) measured at various temperatures. The MR is negative at 2 K with -30 % at 9 T. The negative MR is indeed one of the strong signatures for VRH type transport.[16, 17, 18] Further, the negative MR diminishes with increase in temperature and turns to positive MR for higher temperatures. The negative MR reaches a minimum at about 20 K which is exactly the temperature range where the changes in trend are simultaneously seen in $R_s$ and $n_s$. The origin of negative MR in VRH is attributed to the disruption of the interference effects between forward scattering events in the presence of magnetic field. By considering the interference among the hopping paths between two sites at a distance $R_M$ apart, where $R_M$ is the optimum hopping length, it is shown that this interference affects the hopping probability between these two sites. Inset of Fig. 4a shows the temperature dependence of MR in the negative MR regime at 9 T which scales approximately as $T^{-1}$ as expected in the case of 2D VRH.[19] The temperature dependence of hopping length ($R_M$) basically determines the MR dependence on temperature. Further, the positive MR observed above 20 K arises from Lorentz scattering of electrons in presence of magnetic field, which has been extensively studied for the 2DEG $LaAlO_3$/ $SrTiO_3$ interface.[10, 20, 21] Figure 4b shows the plot of MR as a function of magnetic field, where it shows linear $B$ dependence at high magnetic fields and $B^2$-dependence for low fields (shown in inset), which is in accordance with theory and experiments in VRH regime.[19, 22] To exclude the orbital contributions to MR in VRH regime, measurements were further performed in in-plane mode (current and magnetic field are parallel and in the plane of the 2DEG) which is shown in Fig 4c. In this case, it can be seen that MR is negative for the temperatures below 20 K and diminishes and turns to zero for the temperatures above 15 K. The absence of MR above 20 K suggests the absence of Lorentz scattering of electrons in this in-plane geometry. Figure 4d shows the angle dependence of $R_s$ measured at 2 K and 9 T. Here 0 degree corresponds to in plane mode and 90 degree corresponds to out of-plane mode. It can be observed that the $R_s$ show large anisotropy between in-plane and out of-plane modes, which is due to the confinement effect of the 2DEG. These MR measurements further confirm that the transport mechanism at low temperatures for the thicker samples is governed by VRH and this VRH regime starts to appear about 20 K which is evidenced by the change in the MR (from positive to negative). Generally, the hopping transport favors over other mechanisms when a considerable amount of localization is present in the system, the carrier confinement would also be crucial in these 2D systems where the absence of lateral transport can further favors the hopping transport. Finally, we comment on the unusual carrier density recovery at low temperatures. The coincidence of the critical temperature with the evolution of VRH regime, where the carrier density increases as the temperature decreases suggests that it may relate to hopping transport. Thus we suggest that the overall carrier density behavior could be understood by considering two regimes of transport mechanisms where initially the carrier density decreases with decreasing temperature due to carrier activation and once the a critical lower carrier density is approached then the carrier density starts to increase due to variable range hopping process. The way we can interpret this as in the hopping transport, some of the localized electrons would become electrically mobile in hopping between nearest neighbour sites, thus enhances the mobile carrier density in this regime.

In summary, we have shown the thickness dependence of localization effects of 2DEG at $NdAlO_3/SrTiO_3$ interfaces. The emergence of 2D VRH is observed at low temperatures at about 20 K in thicker samples. The VRH transport mechanism is further confirmed by a detailed study of magneto resistance. Our results further emphasize the dominant role of overlayer thickness in controlling the transport properties of 2DEG at these interfaces through the interface strain effects. Moreover the study presented in this report can certainly assist to distinguish the localization from other mechanisms which is very important in understanding the physics of electron transports in two dimensional oxide systems.

**Acknowledgement**

We thank the National Research Foundation (NRF) Singapore under the Competitive Research Program (CRP) "Tailoring Oxide Electronics by Atomic Control" NRF2008NRF-CRP002-024, National University of Singapore (NUS) cross-faculty grant and FRC (ARF Grant No. R-144-000-278-112) for financial support.

**Figure captions:**

**FIG. 1**. (a) RHEED oscillations obtained during the growth of the $NdAlO_3/SrTiO_3$ (100) samples (b) AFM image of the $NdAlO_3/SrTiO_3$ sample showing the step flow surface indicating the layer by layer growth of the $NdAlO_3$ films.

**FiG. 2**. Temperature dependence of (a) Sheet resistance, $R_s$, (b) Charge carrier density, $n_s$, and mobility, $\mu$, for $NdAlO_3/SrTiO_3$ samples with different thicknesses.

**FIG. 3**. The $\ln(R_s)$ vs. $(1/T)^{1/3}$ graph for 12 and 16 uc $NdAlO_3/SrTiO_3$ samples, a 2D variable range hopping (VRH) fit.

**FIG. 4**. (a) Out-of-plane MR measured at different temperatures for 12 uc $NdAlO_3/SrTiO_3$ sample. Inset: scaling of MR at 9 T with temperature for negative MR part. (b) MR (Out of plane) measured at 2 K with magnetic field showing linear variation at high magnetic fields. Inset: $B^2$ dependence for low magnetic fields. (c) In-plane MR measured at different temperatures. (d) Angle dependence of $R_s$ at 2 K and 9 T.

**Annadi et al., FIG. 1**

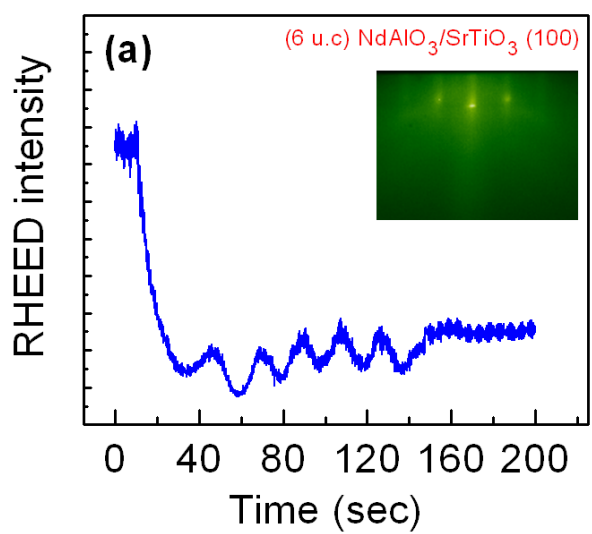 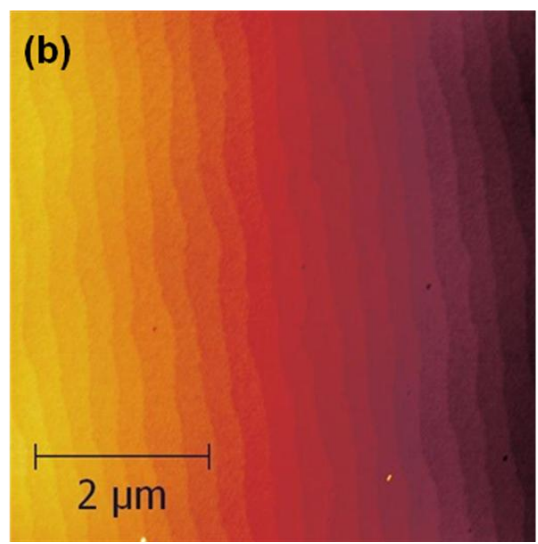

Annadi et al., FIG. 2

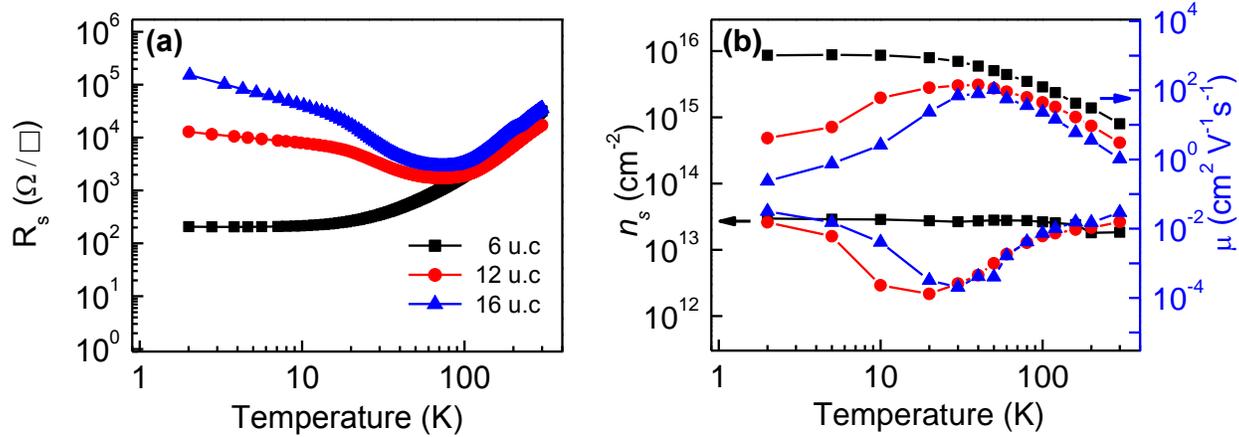

**Annadi et al., FIG. 3**

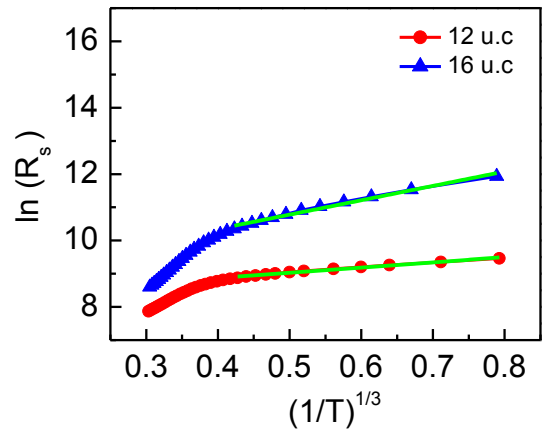

Annadi et al., FIG. 4

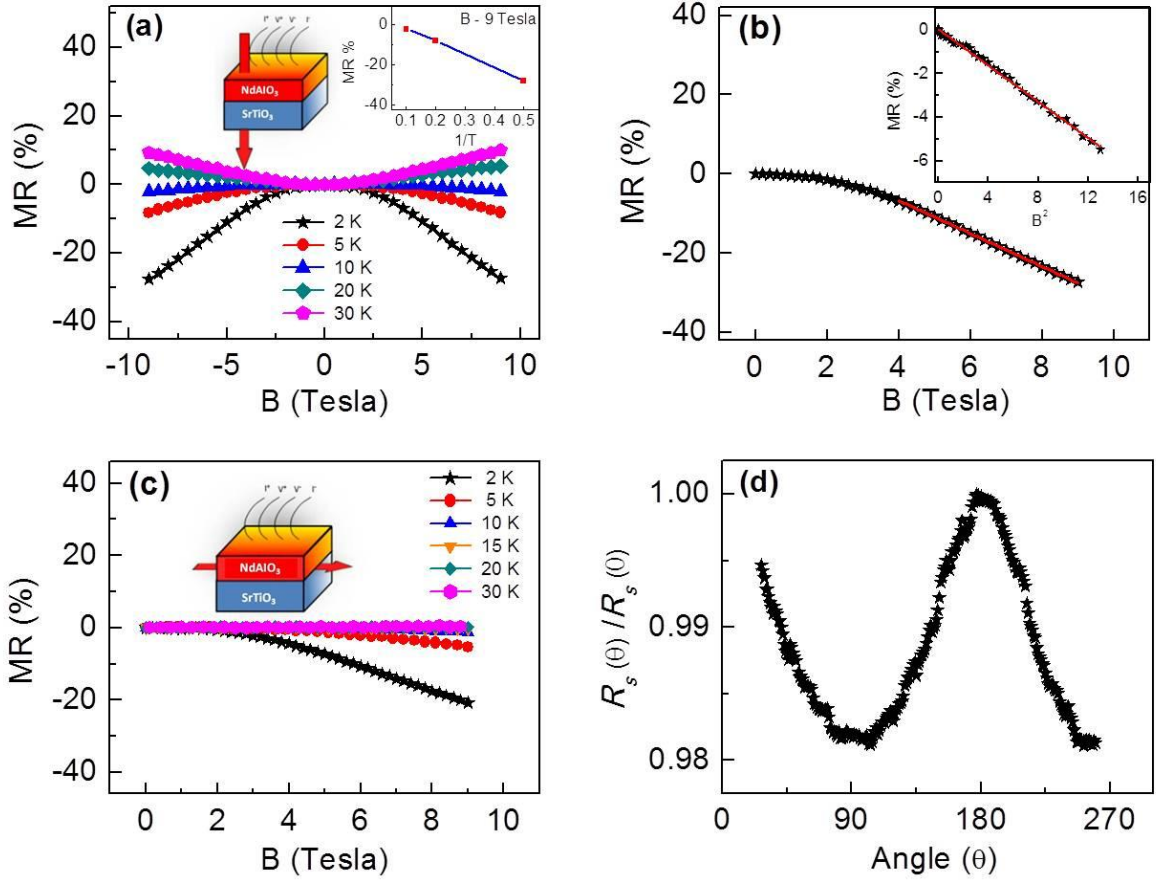